\documentclass[a4paper,11pt]{article}
\pdfoutput=1

\usepackage{amssymb,mathrsfs,amsmath}
\usepackage{a4wide}
\usepackage{color,xcolor}
\usepackage{slashed,soul}
\usepackage{graphicx}
\usepackage{amsfonts}
\usepackage{lscape}
\def\linkcolor{cyan!70!black}

\usepackage[
colorlinks=true
,urlcolor=\linkcolor
,anchorcolor=\linkcolor
,citecolor=\linkcolor
,filecolor=\linkcolor
,linkcolor=\linkcolor
,menucolor=\linkcolor
,linktocpage=true
,pdfproducer=medialab
,pdfa=true
]{hyperref}
\usepackage{amsthm}
\usepackage{booktabs}
\usepackage{array}
\usepackage{rotating}
\usepackage[numbers, sort&compress]{natbib}
\usepackage{multirow}
\usepackage{float}
\usepackage[utf8]{inputenc}
\usepackage[T1]{fontenc}
\usepackage{appendix}
\usepackage{orcidlink}

\newcommand{\beq}{\begin{equation}} 
\newcommand{\eeq}{\end{equation}} 
\newcommand{\ba}{\begin{array}}  
\newcommand{\ea}{\end{array}} 
\newcommand{\bea}{\begin{eqnarray}}  
\newcommand{\eea}{\end{eqnarray} }  
\newcommand{\bal}{\begin{align}}
\newcommand{\eal}{\end{align}}   
\newcommand{\bi}{\begin{itemize}}  
\newcommand{\ei}{\end{itemize}}  
\newcommand{\ben}{\begin{enumerate}}  
\newcommand{\een}{\end{enumerate}}  
\newcommand{\bc}{\begin{center}}
\newcommand{\ec}{\end{center}} 
\newcommand{\bt}{\begin{table}}
\newcommand{\et}{\end{table}}  
\newcommand{\btb}{\begin{tabular}}
\newcommand{\etb}{\end{tabular}}

\renewcommand{\baselinestretch}{1.2}

\let\OLDthebibliography\thebibliography
\renewcommand\thebibliography[1]{
  \OLDthebibliography{#1}
  \setlength{\parskip}{0pt}
  \setlength{\itemsep}{0pt plus 0.3ex}
}

\allowdisplaybreaks

\begin{document}

\vspace{1cm}

\begin{titlepage}

\vspace*{-1.0truecm}
\begin{flushright}
IFT-UAM/CSIC-23-49
 \end{flushright}
\vspace{0.8truecm}

\begin{center}
\renewcommand{\baselinestretch}{1.8}\normalsize
\boldmath
{\LARGE\textbf{
Lepton Flavor Violation from diphoton effective interactions
}}
\unboldmath
\end{center}

\vspace{0.4truecm}

\renewcommand*{\thefootnote}{\fnsymbol{footnote}}

\begin{center}

{\bf Fabiola Fortuna$\,^a$\footnote{\href{mailto:fabiola.fortuna@cinvestav.mx}{fabiola.fortuna@cinvestav.mx}}\orcidlink{0000-0002-8938-7613}, 
Xabier Marcano$\,^b$\footnote{\href{mailto:xabier.marcano@uam.es}{xabier.marcano@uam.es}}\orcidlink{0000-0003-0033-0504},
Marcela Mar\'in$\,^a$\footnote{\href{mailto:bibiana.marin@cinvestav.mx}{bibiana.marin@cinvestav.mx}}\orcidlink{0000-0001-8520-6582},
and Pablo Roig$\,^{a}$\footnote{\href{mailto:pablo.roig@cinvestav.mx}{pablo.roig@cinvestav.mx}}\orcidlink{0000-0002-6612-7157}}
\vspace{0.5truecm}

{\footnotesize

$^a${\sl Departamento de F\'isica, Centro de Investigaci\'on y de Estudios Avanzados del Instituto Polit\'ecnico Nacional, Apdo. Postal 14-740, 07000 Ciudad de M\'exico, M\'exico. \vspace{0.15truecm}}

$^b${\sl  Departamento de F\'{\i}sica Te\'orica and Instituto de F\'{\i}sica Te\'orica UAM/CSIC,\\[-.7ex]
Universidad Aut\'onoma de Madrid, Cantoblanco, 28049 Madrid, Spain \vspace{0.15truecm}}
}

\end{center}

\renewcommand*{\thefootnote}{\arabic{footnote}}
\setcounter{footnote}{0}

\vspace{0.4cm}
\begin{abstract}
\noindent 
We consider charged lepton flavor violating transitions mediated by the diphoton effective interactions $\ell_i\ell_j\gamma\gamma$ and explore which processes can probe them better. 
Our analysis includes single and double radiative decays, $\ell_i\to\ell_j\gamma(\gamma)$, as well as $\ell_i\to\ell_j$ conversions in nuclei for all possible flavor combinations, which we compute for the first time for $\ell\to\tau$ conversions in this framework.
We find that currently the best limits are provided by the loop-induced $\ell_i\to\ell_j\gamma$ processes, while the best future sensitivities come from $\mu\to e$ conversion in aluminum and from potential $\tau\to \ell\gamma\gamma$ searches at Belle II or at the Super Tau Charm Facility.  
We also motivate the search for $\mu\to e\gamma\gamma$ at the Mu3e experiment as a complementary probe of these operators.

\end{abstract}

\end{titlepage}


%
\section{Introduction}

The discovery of neutrino oscillations~\cite{Super-Kamiokande:1998kpq, SNO:2001kpb, SNO:2002tuh} implies the non-conservation of  lepton flavor and it calls for an extension of the Standard Model to accommodate non-vanishing neutrino masses.
Although no other lepton flavor violating process has been observed so far, it is natural to query if the dynamics behind neutrino oscillations can produce charged lepton flavor violating processes~\cite{Masiero:2004js, Mihara:2013zna, Calibbi:2017uvl}.

Experimental searches for charged lepton flavor violation (cLFV)  date back to the late 1940’s, and the upper limits have improved ever since until the impressive current bounds of $\mathcal O(10^{-13})$ for some $\mu\to e$ transitions, with an even more promising future landscape \cite{Lee:2022moh, Davidson:2022jai}. 
cLFV processes involving taus imply a greater experimental challenge~\cite{Marciano:2008zz, Bernstein:2013hba, Banerjee:2022xuw, Davidson:2022jai}, but the strong experimental effort has pushed current upper limits to the $\mathcal O(10^{-8})$ level, which will be further improved in the future~\cite{Banerjee:2022xuw} at Belle~II~\cite{Belle-II:2018jsg}, the Super Tau Charm Factory (STCF)~\cite{Achasov:2023gey}, or the Future Circular Collider~\cite{FCC:2018byv, FCC:2018evy, FCC:2018vvp}. 
We summarize in Table~\ref{tab:BRsExp} the current and future status of some cLFV processes, the most relevant ones for our analysis.

In the literature, we find plenty of models proposed to describe  cLFV interactions. 
The most popular cLFV processes, both from the experimental and theoretical sides, are probably the radiative  $\ell_i\to\ell_j\gamma$ decays.
Nevertheless, if cLFV was discovered, experimental input on a multitude of independent processes would assist in discriminating among these UV completions, which introduces the need to explore also other cLFV transitions. 
In this work, we will focus on $\ell_i\to\ell_j$ conversion in nuclei and on the double radiative $\ell_i\to\ell_j\gamma\gamma$ decays.

The $\ell_i\to\ell_j$ conversions in nuclei are well-motivated scenarios to study cLFV interactions. 
The $\mu\to e$ conversion in nuclei has already been pursued in the past, with the strongest limit given by Sindrum~II~\cite{SINDRUMII:2006dvw}.
Nuclei transitions involving $\tau$ leptons are a bit different since they actually refer to the conversion of an electron or a muon into a tau via deep inelastic scattering (DIS) with a nucleus~\cite{Gninenko:2001id, Sher:2003vi}.
The relevant quantity in this case is given by the ratio of the cross-sections of two inclusive processes:
\begin{equation}\label{eq:NA64sensitivity}
 \mathcal{R}_{\tau\ell}=\frac{\sigma(\ell \mathcal{N}\to\tau X)}{\sigma(\ell \mathcal{N}\to\ell X)}\,,
\end{equation}
where the denominator is given by the dominant contribution to the inclusive $\ell+\mathcal{N}$ process as a result of the lepton bremsstrahlung on nuclei~\cite{Gninenko:2018num}.
At present, there are no experimental limits on these transitions, however, the NA64 experiment at the CERN SPS plans to search for them~\cite{Gninenko:2018num}, with an expected sensitivity of $\mathcal{R}_{\tau\ell}\sim[10^{-13},10^{-12}]$.
In addition to the NA64, future foreseen  experiments such as the muon collider~\cite{Delahaye:2013jla}, the electron-ion collider (EIC)~\cite{Deshpande:2013paa}, the international linear collider (ILC)~\cite{Baer:2013cma} or circular colliders as LHeC~\cite{Acar:2016rde} might search for this conversion.
All these searches for $\ell\to\tau$ transitions will have a relevant impact on different models or effective approaches~\cite{Gninenko:2001id, Sher:2003vi, Kanemura:2004jt, Abada:2016vzu, Takeuchi:2017btl, Husek:2020fru, Antusch:2020vul, Cirigliano:2021img, Ramirez:2022zpk}.

\begin{table}[t!]
\begin{center}
\setlength{\tabcolsep}{6pt}
\begin{tabular}{lllll}
\hline
\hline
cLFV obs. & \multicolumn{2}{l}{Current upper limit  (90\%CL)} & \multicolumn{2}{l}{Expected future limits }\\
\hline
$\mu\to e\gamma$ &  $4.2\times10^{-13}$ & MEG (2016)~\cite{TheMEG:2016wtm}& $6\times10^{-14}$ & MEG-II~\cite{MEGII:2018kmf} \\
$\mu\to e\gamma\gamma$ &  $7.2\times10^{-11}$ & Crystal Box (1986)~\cite{Grosnick:1986pr}& --- & --- \\
$\mu A\to eA$ &  $7\times10^{-13}$ & Sindrum II (2006)~\cite{SINDRUMII:2006dvw}& $6.2\times10^{-16}$ & Mu2e~\cite{Mu2e:2022ggl}\\
$\tau\to e\gamma$ &  $3.3\times10^{-8}$ &BaBar (2010)~\cite{Aubert:2009ag}& $9\times10^{-9}$ & Belle-II~\cite{Banerjee:2022xuw} \\
$\tau\to e\gamma\gamma$ &  $2.5\times10^{-4}$ & Bryman {\it et al.} (2021)~\cite{Bryman:2021ilc}& --- &--- \\
$\tau\to \mu\gamma$ &  $4. 2\times10^{-8}$ &Belle (2021)~\cite{Belle:2021ysv}& $6.9\times10^{-9}$& Belle-II~\cite{Banerjee:2022xuw} \\
$\tau\to \mu\gamma\gamma$ &  $1.5\times10^{-4}$ & ATLAS (2017)~\cite{Angelozzi:2017oeg}& --- &--- \\
$\ell \mathcal{N}\to \tau X$ & 
    --- & --- & $[10^{-13},10^{-12}]$ & NA64~\cite{Gninenko:2018num} \\
\hline
\hline
\end{tabular}
 \caption{Experimental upper bounds and future expected sensitivities for the set of cLFV transitions relevant to our analysis. In the process $\mu A\to eA$ conversion, A=Au for the current upper limit,  while A=Al for the expected future limit. 
 Note there are also promising sensitivities at STCF for LFV $\tau$ decays~\cite{Achasov:2023gey}, nevertheless, we will mostly use Belle-II numbers for the future, as they are expected to be released sooner.} \label{tab:BRsExp}  
\end{center}
\end{table}

On the other hand, the $\ell_i\to\ell_j\gamma\gamma$ decays are a good example of less considered but interesting processes~\cite{Bowman:1978kz, Gemintern:2003gd,Cordero-Cid:2005vca, Aranda:2008si, Aranda:2009kz,Davidson:2020ord,Bryman:2021ilc,Fortuna:2022sxt}.
Compared to $\ell_i\to\ell_j\gamma$, the double photon emission has the advantage of not being chirally suppressed and, from an effective field theory (EFT) point of view, it
is sensitive to a different set of operators. 
Experimentally, $\mu\to e\gamma\gamma$ has been searched for by several experiments, and the best upper limits are still provided by the Crystal Box detector~\cite{Grosnick:1986pr}.
The $\tau\to\mu\gamma\gamma$ decays have rarely been searched for, with the only existing direct experimental search being performed by ATLAS~\cite{Angelozzi:2017oeg}.  
No direct search for $\tau\to e\gamma\gamma$ has been carried out. Nevertheless, experimental upper bounds can be obtained by recasting the searches for $\tau\to e\gamma$~\cite{Bryman:2021ilc}.
Moreover, following an EFT analysis, stronger upper limits can be derived, as we recently shown in Ref.~\cite{Fortuna:2022sxt} and further detailed in this work.
For the future, we are unaware of any concrete plans for new $\ell_i\to\ell_j\gamma\gamma$ searches. Nonetheless, they could --in principle-- be searched for at future facilities such as Mu3e~\cite{Blondel:2013ia} for $\mu\to e\gamma\gamma$ and Belle II or the STCF for $\tau\to\ell\gamma\gamma$.
Since they are 3-body decays, and given the experimental similarities between photons and electrons, we could naively expect similar sensitivities than for $\mu\to eee$ and $\tau\to eee$, which are expected to reach the $10^{-16}$ and $10^{-9}$ level\footnote{Furthermore, the prospects for STCF with 10~ab$^{-1}$ reach sensitivities of $\mathcal O(10^{-10})$ for LFV $\tau$ decays~\cite{Achasov:2023gey}.}, respectively.

Motivated by these cLFV processes, we are interested in the less commonly considered contact interactions involving two leptons of different flavor and two photons~\cite{Bowman:1978kz}. 
Such interactions could mediate various cLFV processes, in particular, the $\ell_i\to\ell_j\gamma(\gamma)$ decays and $\ell_i\to\ell_j$ conversion in nuclei, as has been studied in the literature.
In particular, Davidson {\it et al.}~\cite{Davidson:2020ord} have addressed the role of these interactions in both $\mu\to e\gamma\gamma$ and $\mu\to e$ conversion in nuclei\footnote{See also, {\it e.g.}~Refs.~\cite{Kitano:2002mt, Deppisch:2005zm, Cirigliano:2009bz, Dinh:2012bp, Alonso:2012ji, Crivellin:2017rmk}.}, showing that the latter can probe these diphoton operators more efficiently than the former.
It is natural then to wonder if the same is true for $\tau-\ell$ transitions, although no such study exists. 
On the other hand, and as already mentioned, loop-induced $\ell_i\to\ell_j\gamma$ transitions have been computed in Ref.~\cite{Fortuna:2022sxt}, showing a much better sensitivity for these diphoton interactions. 

Following all this previous discussion, our goal is to perform a systematic study of the reach of each of these kinds of processes to probe the diphoton cLFV interactions.
More precisely, we would extend previous works by first computing the $\ell\to\tau$ conversion in nuclei mediated  by these diphoton interactions, to then conclude on the most promising observables both at the current and future experimental landscape.

The paper is organized as follows: in section~\ref{Sec:Observables} we present the diphoton  effective operators that generate cLFV and describe the processes mediated by them.
This includes in particular our new results for $\ell\to \tau$ conversion in nuclei, although some details are left for Appendix \ref{App:A}.
Section~\ref{Sec:Results} is devoted to our numerical analysis and the discussion about the best observable to study the diphoton cLFV operators. 
Finally, we conclude in section~\ref{Sec:Conclusions}.

\section{cLFV form diphoton effective interactions}
\label{Sec:Observables}

We are interested in the less explored local interactions between two charged leptons of different flavor and two photons. 
The lowest-dimension low-energy effective Lagrangian that describes the local interaction $\bar{\ell}_{i}\ell_j\gamma\gamma$, where  $\ell_i,\ell_j=e,\mu,\tau$, has  energy dimension 7 and is given by~\cite{Bowman:1978kz} 
\begin{equation} \label{eq:Leff}
  \mathcal{L}_\text{eff}=\left(G_{SR}^{\, ij}\bar{\ell}_{L_i}\ell_{R_j}+G_{SL}^{\, ij}\bar{\ell}_{R_i}\ell_{L_j}\right)F_{\mu\nu} F^{\mu\nu} 
+\left(\tilde{G}_{SR}^{\,ij}\bar{\ell}_{L_i}\ell_{R_j}+\tilde{G}_{SL}^{\,ij}\bar{\ell}_{R_i}\ell_{L_j}\right)\tilde{F}_{\mu\nu}F^{\mu\nu} +
{\rm h. c.}\,,
\end{equation}
where the subscripts $L(R)$  stand for the left(right)-handed chirality of the lepton and $\tilde{F}_{\mu\nu}=\frac{1}{2}\epsilon_{\mu\nu\sigma\lambda}F^{\sigma\lambda}$ is the dual electromagnetic field-strength tensor. 
Notice that Ref.~\cite{Bowman:1978kz} also considered the dim-8 operator $\bar{\ell}_{L_i}\gamma^\sigma\ell_{L_j}F^{\mu\nu}\partial_\nu F_{\mu\sigma}$, as well as the analogous operators for the right-handed fermions.
Nevertheless, we do not include their contribution since it is suppressed by higher powers of the cut-off scale of the EFT.

This kind of interaction leads to cLFV processes such as $\ell_i\to\ell_j\gamma(\gamma)$ and $\ell_i\to\ell_j$ conversion in nuclei, as we detail in the following.
They also lead to other cLFV processes including the 3-body $\ell_i\to\ell_j\bar{\ell_k}\ell_k$ decays or semileptonic tau decays $\tau^-\to\ell^- P P$, where $PP$ are a pair of light pseudoscalar mesons (which could be detected as a vector resonance).
Nevertheless, we do not expect them to be competitive with respect to the other decays in restricting the couplings of Eq.~(\ref{eq:Leff}), and
we will therefore not discuss them here.

Notice that these processes involving two photons can also be induced by the dim-5 interaction involving just one photon, with the second photon being emitted from either of the charged particles in the process. 
Nevertheless, these dim-5 interactions are already severely constrained by $\ell_i\to\ell_j\gamma$ and we do not expect to learn anything new by considering the $\alpha$-suppressed processes with two photons (see for instance Ref.~\cite{Fortuna:2022sxt} for a more quantitative discussion).
For all these reasons, in the following, we will focus only on the phenomenology associated with the dim-7 Lagrangian in Eq.~\eqref{eq:Leff}.

\subsection[$\ell_i\to\ell_j\gamma\gamma$]{$\boldsymbol{ \ell_i\to\ell_j\gamma\gamma}$}
\label{sec:liljgg}

At tree level, the  effective Lagrangian in Eq.~(\ref{eq:Leff}) generates the decay $\ell_i\to\ell_j\gamma\gamma$, as it was first computed by Bowman {\it et al}~\cite{Bowman:1978kz}.
In our analysis, we will only need the total decay rate, which is given by
\begin{equation} \label{eq:rate_2gamma}
\Gamma(\ell_i\to\ell_j\gamma\gamma) =  \frac{|G_{ij}|^2 }{3840 \pi ^3}\, m_i^7 \,,
\end{equation}
where we have neglected the mass of the final lepton and $G_{ij}$ is defined as  
\begin{equation} \label{eq:Gij}
	|G_{ij}|^2 = |G_{SL}^{\,ij}|^2 + |G_{SR}^{\,ij}|^2 + |\tilde G_{SL}^{\,ij}|^2 + |\tilde G_{SR}^{\,ij}|^2.
\end{equation}

Note that the rate in Eq.~\eqref{eq:rate_2gamma} seems to imply a non-decoupling behavior, as it appears to be independent of the new physics.
The reason is that the effective couplings in $G_{ij}$ are defined as dimensionful couplings, as can be seen in the Lagrangian in Eq.~\eqref{eq:Leff}. Therefore, they have an implicit dependence on the cut-off scale of $1/\Lambda^3$, which ensures the decoupling behavior of this and the rest of the transition rates explored in this work.

\subsection[$\ell_i\to\ell_j\gamma$]{$\boldsymbol{\ell_i\to\ell_j\gamma}$}

\begin{figure}[t!]
\begin{center}
    \includegraphics[width=.95\textwidth]{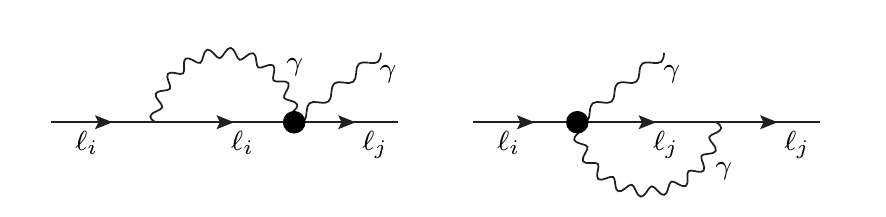}
    \caption{One-loop diagrams generating $\ell_i\to\ell_j\gamma$ from the effective dim-7 diphoton operators in Eq.~\eqref{eq:Leff}, represented as black circles.}\label{fig:loops}
\end{center}
\end{figure}

At one-loop level, the Lagrangian in Eq.~(\ref{eq:Leff}) also generates the single photon decay $\ell_i\to\ell_j\gamma$.
The diagrams, obtained after closing the loop with one of the photons, are shown in Fig.~\ref{fig:loops} and were computed first in Ref.~\cite{Fortuna:2022sxt}.
For our numerical analysis, it will be enough to consider only the leading-log contributions to the decay rate, which are given by
\begin{equation}\label{eq:rate_gamma}
 \Gamma(\ell_i\to\ell_j\gamma)\sim \frac{\alpha |G_{ij}|^2}{256\pi^4}m_{i}^7\log^2\left(\frac{\Lambda^2}{m_i^2}\right)\,, 
\end{equation}
where the mass of the final lepton has been again neglected, $\Lambda$ is the cut-off energy scale associated with our EFT framework, and $G_{ij}$ is defined in Eq.~(\ref{eq:Gij}).
Notice again the apparent non-decoupling behavior, this time log-enhanced, which is again compensated by the $1/\Lambda^3$ scaling of the dimensionful couplings $G_{ij}$.

\subsection[$\ell_i\to\ell_j$ conversion in nuclei]{$\boldsymbol{\ell_i\to\ell_j}$ conversion in nuclei}

As shown by Davidson {\it et al.}, the effective cLFV interaction with two photons also leads to relevant contributions to $\mu\to e$ conversion in nuclei. 
The computation is explained minutely in Ref.~\cite{Davidson:2020ord}, so we only collect the relevant contributions here. 
On the other hand, we extend this idea to flavor transitions with taus, computing the contributions from the Lagrangian in Eq.~\eqref{eq:Leff} to $\ell\to\tau$ conversion in nuclei. 

There are two main contributions to the $\mu\to e$ conversion in nuclei~\cite{Davidson:2020ord}. 
One is the interaction of the leptons with the classical electromagnetic field, and the other is a short-distance loop interaction of two photons with individual protons. 
The former arises from a contact $\mu e\gamma\gamma$ interaction at momentum transfers of order $m_\mu$. 
The latter stems from the loop mixing of the $\bar{e}\mu F_{\mu\nu}F^{\mu\nu}$ operator into the scalar proton operator  $(\bar{e}P_X\mu)(\bar{p}p)$ ($X=L/R$), where overlap integrals, energy ratios, and numerical factors overcompensate for the naive expectation of a loop suppression.
Furthermore, the operators with $F_{\mu\nu}\tilde{F}^{\mu\nu}$ in Eq.~\eqref{eq:Leff} are proportional to $\vec{E}\cdot\vec{B}$, which is negligibly small in the nucleus and can be therefore disregarded.
Then, the conversion rate (CR) in a nucleus $A$ is given by~\cite{Davidson:2020ord}
\begin{equation} \label{eq:mueG}
 {\rm CR}(\mu A\to eA) =  \frac{4m_\mu^5}{\Gamma_{\rm cap}} |\hat G_{\mu e}|^2 \left\vert m_\mu F_A + \frac{18\,\alpha\, m_p}{\pi} S_A^{(p)} \right\vert^2\,,
\end{equation}
where $\Gamma_{\rm cap}$ is the muon capture rate on nucleus $A$ \cite{Suzuki:1987jf} and  $F_A/S_A^{(p)}$ are overlap integrals that can be found in Refs.~\cite{Davidson:2020ord} and~\cite{Kitano:2002mt}, respectively.
Notice that we defined a new effective coupling $|\hat G_{\mu e}|^2\equiv|G_{SL}^{\mu e}|^2+|G_{SR}^{\mu e}|^2$ to emphasize that the contribution from $F_{\mu\nu}\tilde{F}^{\mu\nu}$ terms is negligible. 

The $\mu\to e$ conversion experiments are typically low-energy processes where the muon becomes bounded before decaying in orbit or being captured by the nucleus~\cite{Papoulias:2013gha}. By contrast, the $\ell\to\tau$ conversion experiments are based on a fixed-target nucleus hit by an incoming electron or muon beam. The conversion is expected to occur by DIS of the lepton off the nucleus, meaning that the energy is high enough as to break the nucleons within the nucleus and interact with its partons, {\it i.e.}, quarks and gluons \cite{Gninenko:2018num}. 
Therefore, we focus on the inclusive process, whose products of interaction are a $\tau$ lepton plus any hadrons, {\it i.e.}, $\ell+\mathcal{N} (A, Z)\to \tau + X$, where we do not have any information about $X$.

Husek {\it et al.} \cite{Husek:2020fru} performed an analysis of LFV tau decays and $\ell\to\tau$ conversion in nuclei using SMEFT operators up to dimension 6. 
Here we follow the same methodology to compute the total cross-section in the process $\ell+\mathcal{N} \to \tau + X$ coming from the Lagrangian in Eq.~\eqref{eq:Leff}.

The dynamics of the interacting parton living in the hadronic environment of the nucleus is influenced by low-energy non-perturbative QCD effects. Given that the perturbative cross-sections  are calculated within the framework of perturbation theory, this computation includes a characteristic energy scale, $Q^2$, at which perturbative and non-perturbative effects are factorized. 
The non-perturbative behavior is encoded in the so-called parton distribution functions (PDFs). 
Both the PDFs and the perturbative cross-sections are functions of $Q^2$, typically related to the transferred momentum  $q^2$ of the system as $Q^2=-q^2$. 
In addition, the PDFs are also characterized through the fraction of the nucleus momentum carried by the interacting parton, $\xi$. Therefore, we express the perturbative cross-section as well as the  non-perturbative PDFs as a function of the two discussed invariant quantities. Using the QCD  factorization theorems, we can obtain the total cross-section of the process by calculating the convolution of the perturbative cross-section ($\hat{\sigma}$) with the non-perturbative PDFs ($f$):
\begin{equation}
	\sigma_{\ell-\tau}=\hat{\sigma}(\xi,Q^2)\otimes f(\xi,Q^2)\,.
\end{equation}

The evolution of the PDFs in terms of $Q^2$ is achieved by using the DGLAP evolution equations~\cite{Gribov:1972ri, Dokshitzer:1977sg, Altarelli:1977zs}, whose dependence on the momentum fraction $\xi$ is completely non-perturbative and has to be extracted from the data. 
Moreover, given that the scattering happens with nucleons bounded in a nucleus, the non-perturbative effects relevant to describe the $\ell\to\tau$ conversion in nuclei are better captured by the nuclear parton distribution functions (nPDFs).
For this computation we use the nCTEQ15-np fit of the nPDFs, provided by the nCTEQ15 project~\cite{Kovarik:2015cma}, and incorporated within the ManeParse Mathematica package~\cite{Clark:2016jgm}.

\begin{figure}[t!]
\centering
\includegraphics[width=51mm]{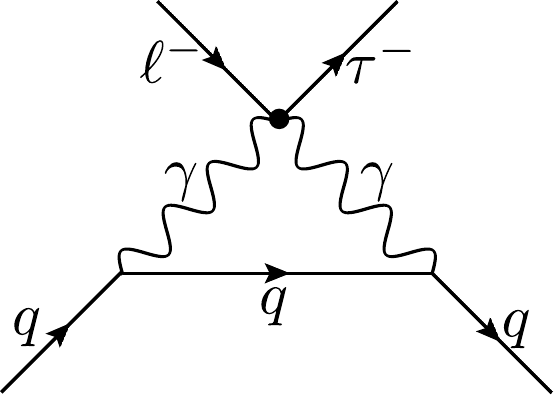} 
\caption{One-loop contribution to $\ell q \rightarrow \tau q$ ($\ell = e, \mu$), stemming from effective $\tau\ell\gamma\gamma$ interactions.}
\label{fig-loop}
\end{figure}

The contributions to the perturbative cross-sections come from the  dimension seven operators in Eq.~(\ref{eq:Leff}). These bring the following contributions in:
\begin{enumerate}
\item The process $\ell q \rightarrow \tau q$ (see Fig.~\ref{fig-loop}) involves a loop with a quark and two photons.
\item The same process as in (a), but with antiquarks: $\ell \bar{q} \rightarrow \tau \bar{q}$. 
Note that the non-perturbative behavior of anti-quarks inside the nucleons differs from their opposite-charged partners, and also the perturbative cross sections of the process are different from those involving quarks.   
\end{enumerate}

The process $\ell g\to\tau g$ also contributes to the perturbative cross-sections, although the operators in Eq.~(\ref{eq:Leff}) generate it at two-loop level. The diagram for this contribution would be similar to the one in Fig.~\ref{fig-loop}, closing the quarks line in a loop and adding initial and final gluons coupled to the  quarks. Given that this process has an additional loop suppression compared with the ones mentioned in (a) and (b),  we neglect it. 

Accounting for  the contributions to the perturbative cross-section previously mentioned,  we obtained the following expression for the unpolarized squared amplitude as a function of $\xi$ and $Q^2$ 
{\small
\begin{align} \label{eq:perturbative}
   \overline{|\mathcal{M}_{qq}(\xi,Q^2)|^2}
   &=~6e^4\Big(|G_{S R}^{\tau\ell}|^2+|G_{SL}^{\tau\ell}|^2\Big)\left[\left(m_\ell^2+m_\tau^2+Q^2\right)\left(\left(m_i+\xi M\right)^2+Q^2\right)\right]\Gamma_{qq}(\xi,Q^2)\nonumber \\
   &+\frac{3e^4}{2}\left(|\tilde{G}_{S R}^{\tau\ell}|^2+|\tilde{G}_{SL}^{\tau\ell}|^2\right)\left[\left(m_\ell^2+m_\tau^2+Q^2\right)\left(\left(m_i-\xi M\right)^2+Q^2\right)\right]  \tilde{\Gamma}_{qq}(\xi,Q^2) \,,
\end{align}
} 

\noindent
where $\Gamma_{qq}(\xi, Q^2)$ and $\tilde{\Gamma}_{qq}(\xi, Q^2)$ result from the evaluation of the loop in Fig.~\ref{fig-loop} and are given in Appendix~\ref{App:A}. 
Notice that the interference between $F_{\mu\nu}F^{\mu\nu}$ and $\tilde{F}_{\mu\nu}F^{\mu\nu}$ term vanishes, while the interference between $L$ and $R$ operators is chirality suppressed and we therefore neglect it.
For the process with antiquarks, we obtain an identical expression as Eq.~(\ref{eq:perturbative}) but with different loop functions (see Appendix~\ref{App:A}).

The perturbative unpolarized differential cross sections can be computed from the squared amplitude in Eq.~(\ref{eq:perturbative}), leading to 
\begin{align}
	\frac{d\hat{\sigma}(\ell\, q_i(\xi P)\rightarrow \tau\, q_i)}{d\xi dQ^2}&=\frac{1}{16\pi \lambda(s(\xi),m_\ell^2,m_i^2)}\overline{|\mathcal{M}_{qq}(\xi,Q^2)|^2}\,,\\
	\frac{d\hat{\sigma}(\ell\, \bar{q}_i(\xi P)\rightarrow \tau\, \bar{q}_i)}{d\xi dQ^2}&=\frac{1}{16\pi \lambda(s(\xi),m_\ell^2,m_i^2)}\overline{|\mathcal{M}_{\bar{q}\bar{q}}(\xi,Q^2)|^2}\,,
\end{align}
with $p_i=\xi P$ the momentum of the interacting parton, $P$ the total momentum of the nucleus and the  subscript $i$ labeling quark flavor. 
We also defined $m_i^2=\xi^2 M^2$, being $M$ the nucleus mass.
$\lambda(s(\xi),m_\ell^2,m_i^2)$  is the usual K\"all\'en  function. Finally, working at leading order in QCD, the total cross-section reads
\begin{align} \label{int-ltau}
	\sigma(\ell \mathcal{N}(P)\rightarrow\tau X) = \sum_i \int_{\xi_\text{min}}^1 \int_{Q^2_{-}(\xi)}^{Q^2_{+}(\xi)} d\xi dQ^2 &\left\lbrace \frac{d\hat{\sigma}(\ell \,q_i(\xi P)\rightarrow \tau\, q_i)}{d\xi \, dQ^2} f_{q_i} (\xi,Q^2)\right. \nonumber \\
	&+ \left. \frac{d\hat{\sigma}(\ell\, \bar{q}_i(\xi P)\rightarrow \tau \, \bar{q}_i)}{d\xi dQ^2} f_{\bar{q}_i}(\xi,Q^2) \right\rbrace\,,
\end{align}
being $f_{q_i} (\xi,Q^2)$ and $f_{\bar{q}_i}(\xi,Q^2)$  the quark and antiquark PDFs, respectively. The integration limits can be found in Appendix E of Ref.~\cite{Husek:2020fru}.

\section{Results and discussion}
\label{Sec:Results}

As detailed in Section~\ref{Sec:Observables}, the diphoton effective interactions in Eq.~\eqref{eq:Leff} contribute to $\ell_i\to\ell_j\gamma$, $\ell_i\to\ell_j\gamma\gamma$ and $\ell_i\to\ell_j$ conversion in nuclei.
Therefore, it is interesting to compare the sensitivity of each of these processes as a probe of the effective couplings $G_{ij}$ or, alternatively, of the new physics scale associated with these effective interactions, which we can naively define as $\Lambda=1/\sqrt[3]{|G_{ij}|}$.

The transition rates for (double) radiative decays and $\mu\to e$ conversion depend on the combination of couplings in Eq.~\eqref{eq:Gij} and thus it is straightforward to translate the experimental upper limits in Table~\ref{tab:BRsExp} into upper limits on $G_{ij}$.
In the case of $\ell\to\tau$ conversions, however, the dependence on the effective couplings does not factorize out in the same manner, see Eq.~\eqref{eq:perturbative}, so we need to make some assumption on their structure. 
For concreteness, we will consider three benchmark scenarios in our numerical analysis:
\begin{align} \label{eq:scenarios}
(i)~|G_{\tau\ell}|^2 &= |G_{S R}^{\tau\ell}|^2 + |G_{S L}^{\tau\ell}|^2=|\tilde{G}_{S R}^{\tau\ell}|^2 + |\tilde{G}_{S L}^{\tau\ell}|^2\,,\nonumber\\
(ii)~|G_{\tau\ell}|^2 &= |G_{S R}^{\tau\ell}|^2 + |G_{S L}^{\tau\ell}|^2;~\tilde{G}_{S R}^{\tau\ell} = \tilde{G}_{S L}^{\tau\ell} =0\,,\\
(iii)~|G_{\tau\ell}|^2 &= |\tilde{G}_{S R}^{\tau\ell}|^2 + |\tilde{G}_{S L}^{\tau\ell}|^2;~G_{S R}^{\tau\ell} = G_{S L}^{\tau\ell}=0\,. \nonumber
\end{align}

Furthermore, according to the prospects of the NA64 experiment \cite{Gninenko:2018num}, we use two specific nuclei in our analysis, Fe(56,26) and Pb(208,82),  as well as $E_e=100\,$GeV and $E_\mu=150\,$GeV for the energies of the incident lepton beams. 
We calculated the integral in Eq.~(\ref{int-ltau}) using either light quarks only or light plus heavy quarks. The criterion to choose our results was the following: if the error of the calculation (due to the PDF uncertainties) using light quarks only, was larger than the contribution of the heavy quarks, then we neglected the latter contributions. That is the case for scenarios $(i)$ and $(iii)$. Conversely, when the contribution of the heavy quarks was larger than the computation uncertainty, we included them, as in scenario $(ii)$.

\begin{figure}[t!]
\centering
\includegraphics[width=.8\textwidth]{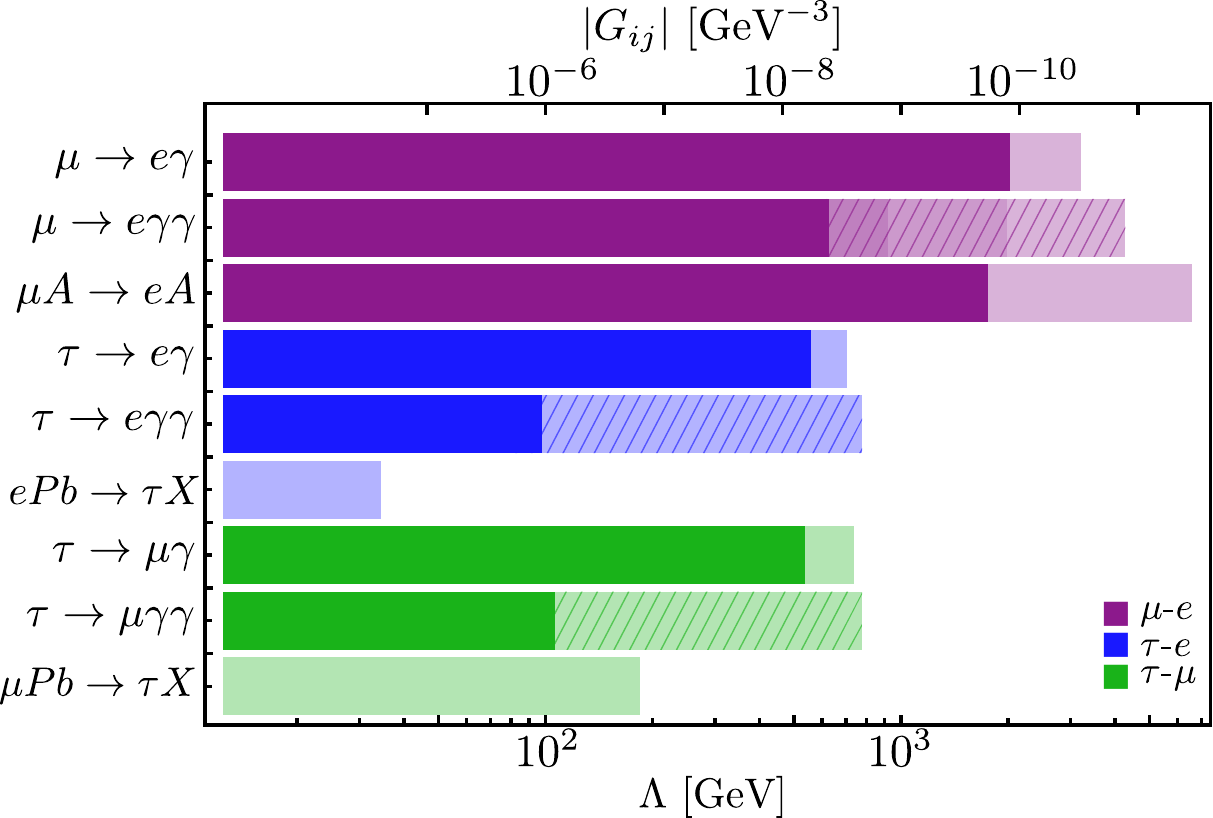}
\caption{Values of the new physics scale $\Lambda$ that are accessible by each of the cLFV observables with current bounds (solid bars) and future sensitivities (lighter bars). 
Striped bars indicate our estimations for future sensitivities in $\ell_i\to\ell_j\gamma\gamma$ decays (see main text for details).
 We define $\Lambda\equiv 1/\sqrt[3]{|G_{ij}|}$.
For $\ell\to\tau$ conversion in nuclei, we used the benchmark (i) in Eq.~\eqref{eq:scenarios}.
}
\label{fig:barras}
\end{figure}

Our results are shown in figure~\ref{fig:barras}, where we present the new physics scale $\Lambda$ accessible by each of the cLFV observables, with solid (light) bars covering current (future) experimental sensitivities.
We see that at present the radiative decays $\ell_i\to\ell_j\gamma$ provide the best sensitivities for $\ell_i\ell_j\gamma\gamma$ effective interactions, as already pointed out in Ref.~\cite{Fortuna:2022sxt}, although $\mu\to e$ conversion is also competitive. 
In fact, future reach in the $\mu\to e$ sector will be dominated by the impressive future sensitivities expected at conversion rate experiments such as Mu2e, going beyond the expected improvement at MEG-II~\footnote{
Additionally, $\mu\to eee$ could also become competitive in the future, since the improved sensitivity of $\mathcal O(10^{-16})$ at Mu3e may overcome the $\alpha$ suppression with respect to $\mu\to e\gamma$. Nevertheless, we expect it to still be below the $\mu\to e$ conversion reach.}. 
On the other hand, and even being a tree-level process in our framework, $\mu\to e\gamma\gamma$ currently provides a much lower sensitivity, since the last search for this process was done by the Crystal Box detector.

We are not aware of any future plan to search for $\mu\to e\gamma\gamma$. Nevertheless, Fig.~\ref{fig:barras} shows how would the future landscape look like assuming an improvement of one, three and five orders of magnitude in this branching ratio.
We see that an improvement of three orders of magnitude is needed to reach the current sensitivity levels of $\mu\to e\gamma$ and $\mu\to e$ conversion, while five orders would make this observable quite competitive with respect to future $\mu\to e$ conversion searches.
Interestingly, the latter corresponds to a sensitivity level of $10^{-16}$, precisely the future goal for $\mu\to eee$, which motivates a dedicated search for $\mu\to e\gamma\gamma$ at Mu3e. 

The situation in the tau sector is different, especially in the future. 
So far, the only direct search for $\tau\to\mu \gamma\gamma$ was performed by ATLAS~\cite{Angelozzi:2017oeg}, although the searches for $\tau\to\ell\gamma$ have been recast into upper limits on $\tau\to\ell\gamma\gamma$~\cite{Bryman:2021ilc}.
Their sensitivity for the $\tau\ell\gamma\gamma$ operator is, however, much lower than current ones from $\tau\to\ell\gamma$, which provides, again, the best probe for $\tau\ell\gamma\gamma$ operators~\cite{Fortuna:2022sxt}.
In the near future, the double photon channels can be searched for at Belle II or at STCF with improved sensitivities. 
In Fig.~\ref{fig:barras} we display the future landscape assuming that Belle II can probe $\tau\to\ell\gamma\gamma$ transitions at the $\mathcal O(10^{-9})$ level, as expected for other cLFV tau decays such as $\tau\to\ell\gamma$ or $\tau\to\ell\ell\ell$.
Then, we see that direct searches for $\tau\to\ell\gamma\gamma$ will improve our current knowledge of $\tau\ell\gamma\gamma$ operators and that they will be competitive with future improvements from $\tau\to\ell\gamma$ decays.
This is an exciting result that motivates dedicated searches for $\tau\to\ell\gamma\gamma$ decays at Belle II and STCF.

\begin{figure}[t!]
\centering
\includegraphics[width=.65\textwidth]{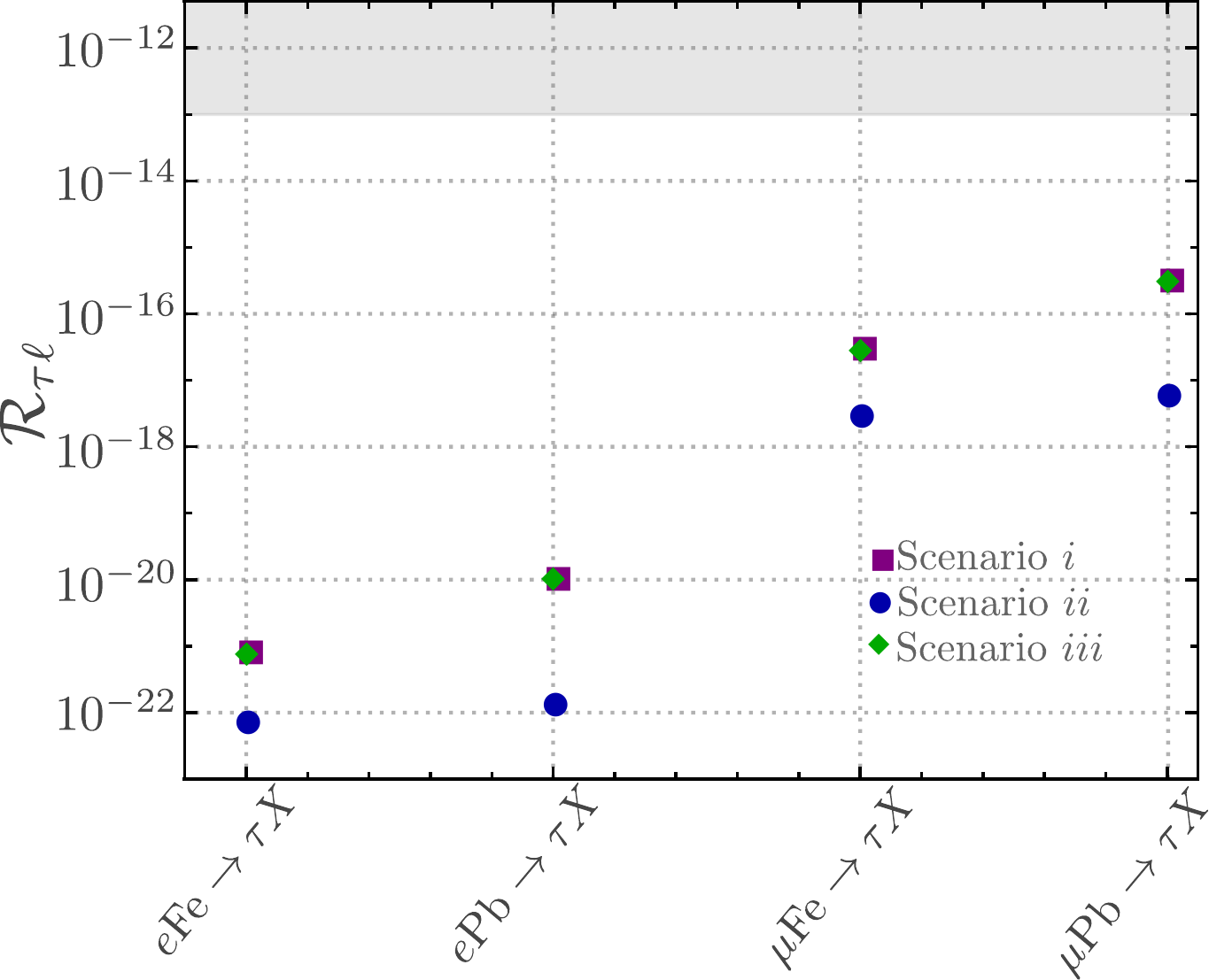} 
\caption{
Maximum $\ell\to\tau$ conversion rates assuming the scenarios in Eq.~\eqref{eq:scenarios} and imposing the upper limits on $\tau\to\ell\gamma$ decays.
The latter is done following Eq.~\eqref{eq:perturbative} with $\Lambda=100$ GeV.
The upper shadowed region shows the expected future sensitivity at NA64.}
\label{fig:results1}
\end{figure}

On the other hand, $\ell\to\tau$ conversion searches at NA64 will not be competitive probing $\tau\ell\gamma\gamma$ operators, especially for $e\to\tau$ conversions.
This is manifest in Fig.~\ref{fig:barras}, which was done considering Pb and scenario $(i)$, and the situation does not improve for other cases we explored, as shown in Fig.~\ref{fig:results1}.
Here, we considered the maximum $G_{\tau\ell}$ allowed by current upper limits on $\tau\to\ell\gamma$ and translated them into $\ell\to\tau$ conversion rates for both Fe and Pb, and for all the scenarios in Eq.~\eqref{eq:scenarios}.
In other words, this figure shows how large the conversion rates can be in each scenario without inducing too large rates for other cLFV transitions. 
We see that the maximum rates are always far from future expected sensitivities at NA64.  
Consequently, and unlike in the $\mu\to e$ sector, the $\ell\to\tau$ conversion does not provide a competitive probe for the local $\tau\ell\gamma\gamma$ interaction.

Finally, a word on the validity of our EFT approach is in order. 
In principle, we only focused on low-energy processes and, therefore, the choice of the low-energy EFT Lagrangian in Eq.~\eqref{eq:Leff} is well justified. 
The only exception could lie in the $\ell\to\tau$ conversion rates, although being a DIS process is not straightforward to conclude it.
The incoming electron (muon) beam has an energy of 100 (150)~GeV, which translates into maximum center-of-mass energy, $\hat s\leq\sqrt{2 M E_\ell}$, around the electroweak scale. 
In that sense, it would be more appropriate to consider a SMEFT approach, where the low-energy diphoton interactions in Eq.~\eqref{eq:Leff} are generated from a $SU(2)_L\times U(1)_Y$ invariant operators. 
Nevertheless, our low-energy analysis showed that the $\ell\to\tau$ conversions are far from being competitive in probing the diphoton operators, which will not change if SMEFT diphoton operators are considered. 

Besides, we see in Fig.~\ref{fig:barras} that the $\Lambda$ scales we are sensitive to are well above the relevant energy scale of each process~\footnote{Again with the only exception of $e\to\tau$ conversion. However, the sensitivity is so far away from other $\tau\to e$ transitions, that the discussion about the validity becomes pointless.}.
In fact, the accessible $\Lambda$ values are in most cases higher than the electroweak scale, motivating a dedicated analysis for diphoton SMEFT operators in cLFV processes, although this goes beyond the scope of our work.

\section{Conclusions}
\label{Sec:Conclusions}

Searches for cLFV processes constitute one of our best prospects to discover new physics, and it is, therefore, important to consider as many cLFV observables as possible. 
This will not only enhance our chances of discovering cLFV transitions, but also provide crucial information about the model behind a potential positive signal, since each BSM scenario predicts in general different correlations between these processes.

In this paper, we have focused on the less studied cLFV diphoton interactions $\ell_i\ell_j\gamma\gamma$ and performed a phenomenological analysis of their implications for the cLFV processes $\ell_i\to\ell_j\gamma$, $\ell_i\to\ell_j\gamma\gamma$, and $\ell_i\to\ell_j$ conversions in nuclei for all possible lepton flavor combinations.
In order to do this, we presented a computation of the contribution to the $\ell\to\tau$ conversion on fixed-target nuclei from the interaction $\tau\ell\gamma\gamma$. 

By comparing with the current experimental upper limits for these processes, we showed that at present the strongest limits for these diphoton operators are provided by the loop-induced $\ell_i\to\ell_j\gamma$ decays, although $\mu\to e$ conversion in nuclei also provides competitive bounds. 
For the future, the strong improvement at Mu2e experiment for $\mu\to e$ conversion in nuclei will dominate the sensitivity for $\mu e\gamma\gamma$ interaction.
This is in contrast to the $\tau$ sector, where future NA64 searches for $\ell\to\tau$ conversion will not be competitive for probing $\tau\ell\gamma\gamma$ operators, and they will be better explored by $\tau\to\ell\gamma$ decays.

Nevertheless, we showed that future direct searches for $\ell_i\to\ell_j\gamma\gamma$ could be competitive or even provide the best limits if they could reach the same sensitivity levels as similar processes such as the 3-body $\ell_i\to \ell_j\ell_j\bar{\ell}_j$ decays.
This result motivates ongoing or planned experiments to search also for the diphoton channel, in particular, $\mu\to e\gamma\gamma$ searches at Mu3e or $\tau\to\ell\gamma\gamma$ at Belle-II, STCF, EIC or FCC.

\paragraph{Acknowledgments.}
We would like to thank Alejandro Ibarra for his collaboration in the first stage of the project and for very valuable discussions.
This project has received support through the Grant IFT Centro de Excelencia Severo Ochoa No CEX2020-001007-S funded by MCIN/AEI/10.13039/501100011033.
FF and MM are indebted to Conacyt for funding. 
XM acknowledges funding from the European Union’s Horizon Europe Programme under the Marie Skłodowska-Curie grant agreement no.~101066105-PheNUmenal. PR is thankful to Cátedras Marcos Moshinsky (Fundación Marcos Moshinsky) and ‘Paradigmas y Controversias de la Ciencia 2022’  (project number 319395, Conacyt) for funding.


\appendix
\section{Functions from the evaluation of the loops in $\boldsymbol{\ell\to\tau}$ conversion in nuclei} \label{App:A}

Here we provide the relevant functions needed to evaluate the perturbative contribution to $\ell\to\tau$ conversion, as given in Eq.~\eqref{eq:perturbative}, which are the result of computing the loop in Fig.~\ref{fig-loop}.
In particular, we define
\begin{equation}
\begin{aligned}
	\Gamma_{qq}(\xi,Q^2)&=\frac{1}{64 \pi^4}|\text{F}1(\xi,Q^2)|^2\,,\\
	\tilde{\Gamma}_{qq}(\xi,Q^2)&=\frac{1}{64 \pi^4} |\text{F}2(\xi,Q^2)|^2\,,\\
	\Gamma_{\bar{q}\bar{q}}(\xi,Q^2)&=\frac{1}{64 \pi^4} |\text{F}3(\xi,Q^2)|^2\,,\\
	\tilde{\Gamma}_{\bar{q}\bar{q}}(\xi,Q^2)&=\frac{1}{64 \pi^4} |\text{F}4(\xi,Q^2)|^2\,,
\end{aligned}
\end{equation}
with
\small{
\begin{align} \label{F1}
	\text{F1} &=2\big[ m(Q^2)+M \xi\big]{\bf B}_0 (M^2 \xi^2; m(Q^2),0)+ 2\big[m(Q^2)+ m_i\big]{\bf B}_0 (m_i^2;m(Q^2),0) \nonumber\\
        &+ 2 \big[m(Q^2) -m_i\big]{\bf B}_0(-Q^2;0,0)+2M \xi \, {\bf B}_1(M^2 \xi^2; m(Q^2),0)+2\big[M\xi-m_i\big] {\bf B}_1(-Q^2; 0,0)\nonumber\\
        &+2m_i\, {\bf B}_1(m_i^2; m(Q^2),0)+2\big[m^3(Q^2) + M m_i\, m(Q^2) \xi- M^2 m_i \xi^2+ m(Q^2)Q^2 - M m_i^2 \xi\big]\nonumber\\ 
        &{\bf C}_0(m_i^2,-Q^2,M^2 \xi^2; m(Q^2),0,0)+2 \big[ m^2(Q^2)-  m_i^2+m_im(Q^2)-m_iM\xi-M^2 \xi^2+M m(Q^2) \xi \big]\nonumber\\
        &\big( M \xi \,{\bf C}_2(m_i^2,-Q^2,M^2 \xi^2; m(Q^2),0,0) +m_i\,{\bf C}_1(m_i^2,-Q^2,M^2 \xi^2; m(Q^2),0,0) \big) \nonumber\\ 
        & +m_i-4m(Q^2)+M\xi\,,
\end{align} 
}
\vspace{-.5cm}
\small{
\begin{align} \label{f2}
	\text{F}2 &= -2 i \bigg( 2\big[ M\xi+ m(Q^2)\big] {\bf B}_0 (M^2\xi^2;m(Q^2),0) + 2 \big[ m_i+m(Q^2)\big] {\bf B}_0 (m_i^2;m(Q^2),0) \nonumber\\
        &+2 \big[M\xi+ m_i-2m(Q^2)\big]{\bf B}_0(-Q^2;0,0)+ 2M\xi\, {\bf B}_1 (M^2\xi^2;m(Q^2),0) +2 m_i\, {\bf B}_1(m_i^2;m(Q^2),0) \nonumber\\	
	&+2\big[m_im^2(Q^2)-2m^3(Q^2)+M m^2(Q^2) \xi-Mm_i^2\xi -M^2m_i\xi^2+2Mm_im(Q^2)+m(Q^2)Q^2\big]\nonumber\\
        &{\bf C}_0 (m_i^2,-Q^2,M^2\xi^2;m(Q^2),0,0)+2\big[m_i^2 -2m_im(Q^2)-M^2\xi^2+2Mm(Q^2)\xi\big]\nonumber\\
        &\big(m_i\,  {\bf C}_1 (m_i^2,-Q^2,M^2\xi^2;m(Q^2),0,0)-M\xi\,{\bf C}_2 (m_i^2,-Q^2,M^2\xi^2;m(Q^2),0,0)\big)\nonumber\\ 
	& -3 (m_i+ M\xi)\bigg)\,,
\end{align}} 
\vspace{-.5cm}
\small{
\begin{align} \label{f3}
	\text{F}3 &= 2\big[m(Q^2)-M\xi\big]{\bf B}_0 (M^2\xi^2;m(Q^2),0)+2\big[m(Q^2)-m_i\big]{\bf B}_0 (m_i^2;m(Q^2),0)\nonumber\\
        &+2\big[m(Q^2)+m_i\big] {\bf B}_0 (-Q^2;0,0)-2M\xi\,{\bf B}_1(M^2\xi^2;m(Q^2),0)-2m_i\,{\bf B}_1 (m_i^2;m(Q^2),0)\nonumber\\
        &+2\big[m_i-M\xi\big]{\bf B}_1(-Q^2;0,0)+2 \big[ M^2\xi^2 +  m_i \big(M\xi+m(Q^2)\big)+ M m(Q^2)\xi + m_i^2- m^2(Q^2) \big]\nonumber\\ 
        & \big(M\xi\,{\bf C}_2 (m_i^2,-Q^2,M^2\xi^2;m(Q^2),0,0) +m_i\, {\bf C}_1 (m_i^2,-Q^2,M^2 \xi^2; m(Q^2),0,0)\big)\nonumber\\ 
	&+ 2\big[m^3(Q^2)+Mm_im(Q^2)\xi+M^2m_i\xi^2+m(Q^2)Q^2+Mm_i^2\xi\big] {\bf C}_0 (m_i^2,-Q^2,M^2\xi^2;m(Q^2),0,0) \nonumber\\
	&- \big(m_i+4m(Q^2)+M\xi \big)\,,
\end{align}}
\vspace{-.5cm}
\small{
\begin{align} \label{f4}
	\text{F}4 &= 2 i\, \bigg( 2 \big[M\xi- m(Q^2)  \big] {\bf B}_0 (M^2\xi^2; m(Q^2),0)+ 2 \big[ m_i-m(Q^2)\big] {\bf B}_0(m_i^2;m(Q^2),0) \nonumber\\
	&+ 2 \big[m_i+2m(Q^2)+M\xi \big] {\bf B}_0 (-Q^2;0,0) + 2 m_i\, {\bf B}_1 (m_i^2;m(Q^2),0) +2M\xi\, {\bf B}_1 (M^2\xi^2; m(Q^2),0) \nonumber\\
	& + 2\big[ 2m^3(Q^2)+m_im^2(Q^2)+M\xi m^2(Q^2)-M\xi m_i^2-M^2\xi^2m_i-2M\xi m_im(Q^2)-m(Q^2)Q^2\big]  \nonumber\\
	&{\bf C}_0 (m_i^2,-Q^2,M^2 \xi^2;m(Q^2),0,0)+2\big[2m_i m(Q^2)-M^2\xi^2-2M\xi \,m(Q^2)+m_i^2\big] \nonumber\\
	& \big(m_i\,{\bf C}_1 (m_i^2,-Q^2,M^2\xi^2;m(Q^2),0,0)- M\xi\, {\bf C}_2 (m_i^2,-Q^2,M^2\xi^2;m(Q^2),0,0) \big) \nonumber\\
	& -3 (m_i+M\xi) \bigg)\,.  
\end{align}}

\normalsize
The function $m(Q^2)$ represents the running of the quark mass in the loop. For the computation of the quark masses at different energy scales, we use RunDec \cite{Chetyrkin:2000yt}. The notation employed for the Passarino-Veltman loop functions is standard.
We use Package-X \cite{PATEL2015276} to analytically evaluate the loop integrals. CollierLink extends Package-X so that the Passarino-Veltman functions can be directly evaluated, using the COLLIER library \cite{Denner:2016kdg}. Then we use CollierLink to numerically evaluate our expressions.


\bibliographystyle{JHEP} 
\bibliography{biblio-LFV2P}

\end{document}